\documentclass[aps,showpacs,floats, floatfix,amsmath, amssymb,preprint]{revtex4}

\usepackage{latexsym}
\usepackage{amsfonts}
\usepackage{graphicx} 
\usepackage{float}

\begin{document}

\title{Violation of market efficiency in transition economies} \author{
   Boris~Podobnik$^{1}$, Ivo Grosse$^{2}$, Davor~Horvatic$^{3}$, 
  Plamen~Ch Ivanov$^{4}$, Timotej Jagric$^{5}$, \\  and H.E. Stanley$^{4}$}

\affiliation{ $^1$Faculty of Civil Engineering, University of Rijeka,
  Rijeka, Croatia\\
  Zagreb School of Economics and Management, 
  Zagreb, Croatia\\
  $^2$Institute of Plant Genetics and Crop Plant Research (IPK), 06466 Gatersleben,
  Germany\\
  $^3$Faculty of Natural Sciences, University of Zagreb, Croatia \\
  $^3$Center for Polymer Studies and Department of Physics,
  Boston University, Boston,  MA 02215 \\
  $^5$ Faculty of Economics and Business, University of Maribor, Slovenia
  }


\begin{abstract} 
  We analyze the European transition economies and show that time series
  for  most  of major indices exhibit (i) power-law correlations 
  in their values, power-law correlations in their magnitudes,
   and (iii)  asymmetric probability
  distribution. We propose a stochastic model that can 
  generate time series with all the previous features 
  found in the empirical data.
\end{abstract}


\maketitle


 An interesting question in economics is whether markets in transition economies 
 defer in their behavior from developed capital markets. One way to analyze
 possible differences in behavior is to test the weak form of market efficiency
 that  states that the present price of a stock
comprises all of the information about past price values   implying that
stock prices at any future time cannot be predicted. 
In contrast to predominant behavior of financial time series of developed
markets characterized by no or very short serial correlations 
\cite{Fama65,Gra64,Sha77,hiemstra}, it is  believed that
 financial series of
emerging markets exhibit different behavior \cite{jagric}.

 For ten transition 
economies in east and central Europe with statistics reported in Table 1, 
 we analyze  time series of index
returns  $R_t = \log S(t+ \Delta t) - \log S(t)$, 
 daily recorded. 

\begin{figure}
\centerline{
\includegraphics[width=10cm,angle=0]{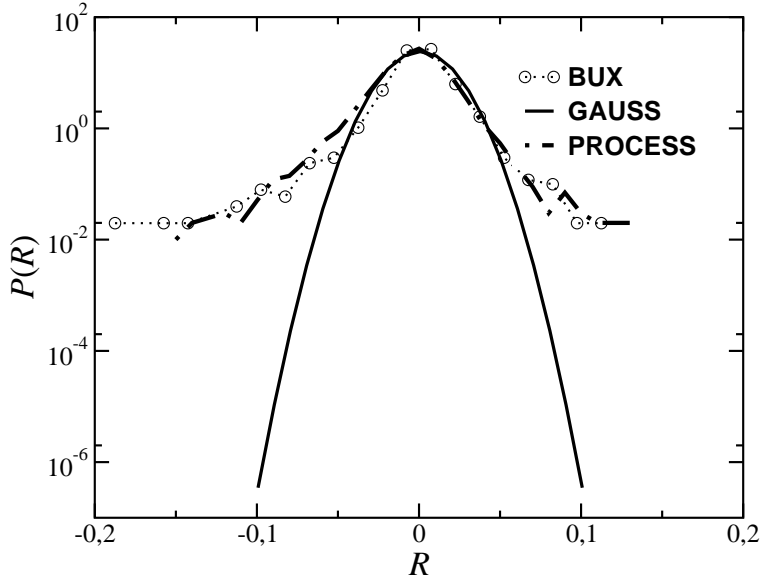}}
\caption{ Probability distribution of  
$R_t$ calculated for  the BUX index and 
the Gaussian distribution with the same standard
  deviation as found for the BUX index. 
  The kurtosis of $P(R_t)$ for
  the BUX index is $17$, which is much greater 
  than the kurtosis  of
   the Gaussian probability distribution, which is $3$. We see that 
  $P(R_t)$ for
  the BUX index is negatively skewed, in opposite to   
  the Gaussian that is symmetric. Shown is $P(R_t)$ of the process with
 $\rho_1=0.09$, $\rho_2=0.3$, and $\lambda = -0.2$}
\label{Fig1}
\end{figure}

Table 1 shows that none of the index time series $R_t$ exhibits a vanishing
skewness defined as a measure of asymmetry ---
 $\langle (x-\mu)^3 \rangle / \sigma^3$ --- where 
$\mu$ and $\sigma$ are the expectation and  the standard deviation,
respectively. Five of time series show  positive
skewness, i.e., their probability distributions have more
 pronounced right tail, 
 while the rest five time series exhibit negative skewness. 
Fig.~\ref{Fig1}  shows the probability distribution 
$P(R_t)$  of the BUX index
with negative skewness and the Gaussian distribution 
clearly with vanishing skewness. 
 
Next we calculate the kurtosis defined as 
$ \langle  (x-\mu)^4 \rangle / \sigma^4$ that is e.g.  for a Gaussian 
distribution  equal to 3. Generally, for a 
probability distribution with more (less) weight in the tails, 
the kurtosis is greater (smaller) than 3. 
 Table 1 shows that for none of the ten index time series 
 the observed probability distribution is a Gaussian. 
 
To analyze correlations in time series, we 
employ the detrended fluctuation analysis (DFA) ~\cite{CKP}, the wavelet
analysis and the Geweke and Porter-Hudak (GPH) method ~\cite{GPH}. 
The detrended fluctuation function $F(n)$ follows a 
scaling law $F(n) \propto n^{\alpha} $
if the time series is power-law auto-correlated.  
A DFA scaling exponent $\alpha > ~ 0.5 $ corresponds to time series 
with power-law correlations, and $\alpha=0.5$ 
corresponds to time series
with no auto-correlations. For  GPH method, the process is said to exhibit
long memory if the  GPH parameter $d$ is from the range
$(0, 0.5)$. 

For each of ten indices time series $R_t$, 
Table 1 shows the DFA scaling exponent $\alpha$, Hurst exponent
$H$ calculated by wavelet analysis, and the  GPH parameter $d$.
We show that DFA and wavelet analysis give similar results.
Besides SAX and perhaps WIG20 index, the other indices exhibit
 power-law serial correlations.  Similar results are 
obtained by GPH method where the relation $\alpha=0.5 + d$ is
expected in presence of
 power-law  correlations. 

Next, we calculate the DFA scaling exponents $\alpha_{|R|}$
 for the time series of $|R_t|$. From 
Table 2 we see that for each index,  the time series $|R_t|$ shows 
power-law auto-correlations, a common behavior on stock markets, 
  where  generally  $\alpha_{|R|} >  \alpha_R$.

In order to investigate to which degree the ten time series exhibit
linear and nonlinear properties \cite{Euro,Ashk20}, we  
phase randomize 
the original time series where the procedure changes (does not change) 
magnitude auto-correlations for a nonlinear (linear)
process \cite{PodHosk}. During phase-randomization procedure
 one performs a Fourier transform of the
original time series and then randomizes the Fourier phases keeping the
Fourier amplitudes unchanged.  At the end, one calculates an inverse Fourier
transform and obtains the surrogate time series $\tilde R_t$.
 
\begin{figure}
\centerline{
\includegraphics[width=10cm,angle=0]{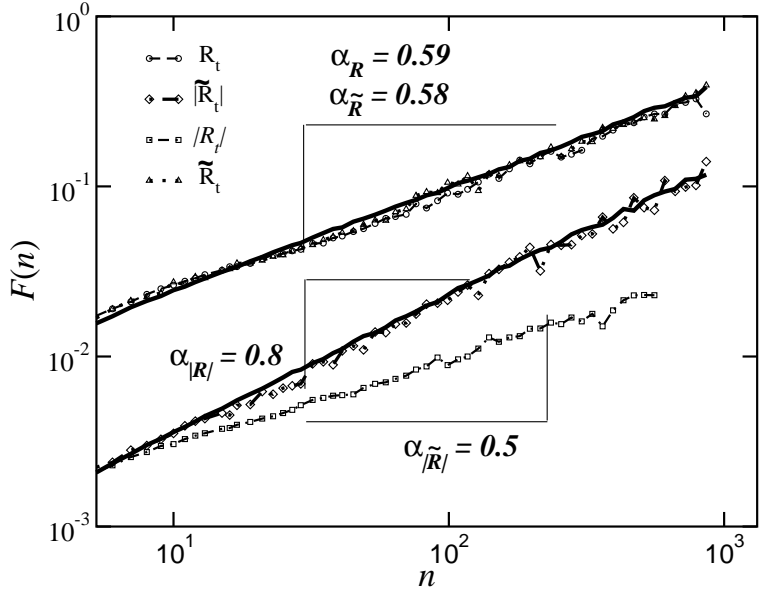}}
\caption{Time series of  returns $R_t$ of Hungarian BUX index.  DFA
  functions calculated for four time series:
    $R_t$, the one obtained after
  phase-randomization procedure $\tilde R_t$, and two magnitudes time series;
  $|R_t|$ and $|\tilde R_t|$. 
 After phase randomization procedure, the time series $|\tilde R_t|$
  has no auto-correlations. By solid lines we show $R_t$
  and $|R_t|$ of the process with $\lambda = -0.2$, $\rho_1=0.09$, and
  $\rho_2=0.3$ }
\label{Fig2}
\end{figure}

 For the BUX index, 
Fig.~\ref{Fig2} shows the DFA functions $F(n)$ of the time
series $R_t$ and  $|R_t|$  together with $F(n)$ of the phase-randomized
surrogate time series $\tilde R_t$ and  $|\tilde R_t|$. As
expected, the $F(n)$ curves of $R_t$
and $\tilde R_t$ are the same~\cite{Euro}. 
In contrast, the  time series $|\tilde R_t|$ is
uncorrelated $(\alpha_{|\tilde R|} = 0.5)$, 
while the time series $|R_t|$ is power-law auto-correlated 
$(\alpha_{|R|} = 0.8)$. Similar behavior in scaling of time series   we find
for all other 10 indices (see Table 1).

Next we propose a stochastic process to model time series  $R_t$
with power-law correlations in both $R_t$ and $|R_t|$
together with  asymmetric 
probability distributions $P(R_t)$ ~\cite{ASYMM}

\begin{eqnarray}
  R_i &= & \sum_{n=1}^{\infty} {a_n(\rho_1) } [
  R_{i-n} -\lambda    |R_{i-n}|  ] +  \sigma_i \eta_i,
\label{mand}\\
\sigma_i &=&  \sum_{n=1}^{\infty} {a_n(\rho_2) } 
\frac{|R_{i-n}|}{  \langle |R_{i}| \rangle }.
\label{granger2}
\end{eqnarray}
The weights defined as
$a_n(\rho) = \rho {\Gamma(n-\rho)} / ({  \Gamma(1 - \rho)  \Gamma(1+n)}) $
for $n>>1$ scales as $a_n(\rho) \propto n^{1-\rho}$, where
$\rho_{1/2} ~ \epsilon ~(0,0.5)$ are scaling parameters. 
 It holds that
$\sum_{n=1}^{\infty} a_n(\rho) =1$.  
If  asymmetry parameter $\lambda$ is zero, 
the process is a combination of two fractionally integrated
processes  in Refs.~\cite{Granger80,Hosk81} and  \cite{Granger96}.
 $\Gamma$ is a Gamma function, and $\eta_i$ denotes
Gaussian white noise with 
 $ \langle \eta_i \rangle = 0$ and  $ \langle \eta_i^2 \rangle =\sigma_0^2$,
 where $\sigma_0^2$ we set to model the variance of empirical data.  

In Ref.~\cite {PodHosk}  for the case $\lambda=0$, $\rho_1 = \rho_2 =\rho$ $(
\rho_{1,2} > 0.5)$, we derived the following two 
scaling relations $\alpha_{R} = 0.5 +
\rho$ and $\alpha_{|R|} = 0.5 + \rho$  between two DFA exponents
$\alpha_{R}$ and $\alpha_{|R|}$ and  $\rho$.  To model empirical time series
with different  exponents $\alpha_{R}$ and $|\alpha_{R}|$, 
we allow  $\rho_1$ and $\rho_2$ to be different.  

Applied the process to model empirical data, first we calculate DFA exponents
$\alpha_R$ and $\alpha_{|R|}$ and if $\alpha_R < \alpha_{|R|}$, 
 we  calculate $\rho_1$ and $\rho_2$ from
scaling relations $\alpha_{R} = 0.5 + \rho_1$ and $\alpha_{|R|} = 0.5 +
\rho_2$, respectively.
For the Hungarian BUX index, from the DFA exponents 
 $\alpha_R = 0.59$ and $\alpha_{|R|} = 0.8$ (see Table 1)   
and previous scaling relations, we calculate  
the parameters $\rho_1 = 0.09$ 
and $\rho_2 = 0.3$. 
In Fig.~2 we show the
scaling function $F(n) \propto n^ {\alpha} $ 
for both model time series $R_t$ and
$|R_t|$ (solid lines), where we arbitrarily set $\lambda = -0.2$ to account for
small skewness in the empirical distribution.
After performing phase-randomization procedure, 
auto-correlations in $|\tilde R_t|$ vanish,
while auto-correlations in $\tilde R_t$ practically  
remain the same as in 
the original time series $R_t$, 
that is the same behavior as we found in empirical data.
 In Fig.1 we also find  that 
$P(R_t)$ calculated for the process 
 fits $P(R_t)$ calculated for
the BUX index.

In conclusion, we  show that for ten transition economies
their market indices analyzed exhibit (i) power-law 
correlations in index returns, (ii) power-law
 correlations in the magnitudes, where 
 the probability distributions exhibit (iii)
asymmetric behavior.  These three properties  we model with 
 a stochastic process specified
by only three parameters.


\begin{table}
\begin{tabular}{ccccccccccc}
\hline
\hline
country & $Rus$ & $Hun$ & $Pol$ &$Slovak$  & $Sloven$ & $Czech$ & $Lit$  &  $Lat$ & $Est$ & $Cro$\\
index & $RTS$ & $BUX$ & $WIG20$ &$SAX$  & $SBI$ & $PX50$ & $VILSE$  &  $RICI$ & $TALSE$ & $CROEMI$\\
st. dev. & $0.031$ & $ 0.017$ & 0.022 & 0.014& 0.014 & $0.014$ & 0.007 &  0.010& 0.019 &   0.014\\
skewness & $-0.344$ & $-0.865$ & -0.446 & -0.409 & 0.416 & $1.342$ & $-1.065$ & 1.240 & 2.944 &  0.731 \\
kurtosis & $7.96$ & $17.69$ & 11.97 & 9.48 & 25.34 & 17.18 & 26.89 & 22.46 & 47.80 & 12.51 \\
$\alpha_{R_t}$          & $0.60$ & $0.59$ & $0.52$ & $0.53$ &  $0.62$ & $0.63$ & $0.63$ & $0.58$ & $0.70$ & 0.58\\
$\alpha_{|R_t|}$        & $0.79$ & $0.80$ & $0.84$ & $0.66$ &  $0.74$ & $0.86$ & $0.69$ & $0.65$ & $0.80$ & 0.70\\
$\alpha_{\tilde R_t}$   & $0.59$ & $0.58$ & $0.52$ & $0.51$ &  $0.58$ & $0.67$ & $0.63$ & $0.56$ & $0.69$ & 0.57\\
$\alpha_{|\tilde R_t|}$ & $0.51$ & $0.50$ & $0.45$ & $0.53$ &  $0.51$ & $0.47$ & $0.54$ & $0.55$ & $0.53$ & 0.52\\
data points & $2232$ & $3373$ & $2530$ & $2204$ &  $2884$ & $2567$ & $1829$ & $2025$ & $183$ &1522\\
\hline
\hline
\end{tabular}
\caption{\label{table1} Basic statistics of financial data. Besides skewness
  and kurtosis, which are the measures for asymmetry and "fatness" in the
  tails, also shown is  DFA
  exponents for time series of indices and their magnitudes together with the
  corresponding values obtained after phase randomization.  }
\end{table}

\begin{table}
\begin{tabular}{ccccccccccc}
\hline
\hline
country & Rus  & Pol & Czech & Hun  &  Slovak & Sloven & Cro &  Lith  &  Latv &  Est \\
index  & RTS & WIG20 & PX50 & BUX & SAX & SBI & CROEMI  & VILSE  & RICI  & TALSE \\ 
$\alpha_R$ & 0.60 & 0.56 &  0.63 & 0.59 &  0.53 & 0.62 & 0.58 & 0.63 & 0.58 & 0.70 \\
$H$  &  0.65 & 0.57 & 0.65 &  0.63  & 0.53 & 0.66 & 0.62 & 0.63 & 0.62 & 0.65\\
$d$  & 0.11 & 0.02  & 0.27 & 0.07  & 0.01 & 0.14 & 0.10 & 0.10 & 0.15 & 0.07\\
\hline
\end{tabular}
\caption{\label{table 2} Scaling exponents calculated for DFA method, wavelet method and GPH method. } 
\end{table}


\end{document}